\documentstyle[multicol,eqsecnum,aps,psfig]{revtex}
\renewcommand{\narrowtext}{\begin{multicols}{2}
\global\columnwidth20.5pc\noindent}
\renewcommand{\widetext}{\end{multicols}
\global\columnwidth42.5pc}
\begin{document}
\draft
\preprint{6 May 2002}
\title{Local excitations in mixed-valence bimetal chains}
\author{Shoji Yamamoto}
\address
{Department of Physics, Okayama University,
 Tsushima, Okayama 700-8530, Japan}
\date{Received 6 May 2002}
\maketitle
\begin{abstract}
We investigate localized intragap states in mixed-valence bimetal
chains in an attempt to encourage further explorations into
quasi-one-dimensional halogen-bridged binuclear metal ($M\!M\!X$)
complexes.
Within a coupled electron-phonon model, soliton and polaron
excitations in the two distinct ground states of $M\!M\!X$ chains are
numerically calculated and compared.
Making a continuum-model analysis as well, we reveal their scaling
properties with particular emphasis on the analogy between $M\!M\!X$
chains and {\it trans}-polyacetylene.
\end{abstract}
\narrowtext

\section{Introduction}

   Halogen ($X$)-bridged transition-metal ($M$) linear-chain
complexes ($M\!X$ chains) \cite{C95} provide a fascinating stage
performed by electron-electron correlation, electron-lattice
interaction, low dimensionality and $d$-$p$ hybridization
\cite{G6408}.
Representative materials such as Wolffram's red
($M=\mbox{Pt},X=\mbox{Cl}$) and Reihlen's green
($M=\mbox{Pt},X=\mbox{Br}$) salts possess mixed-valence ground states
exhibiting intense and dichroic charge-transfer absorption, strong
resonance enhancement of Raman spectra, and luminescence with large
Stokes shift \cite{T343}.
When the platinum ions are replaced by nickel ones, monovalence Mott
insulators are instead stabilized \cite{T4261,T2341}.
Local states such as solitons and polarons \cite{K2122,C723,O2248},
which can be photogenerated or induced by doping, are further topics
of great interest lying in these materials.
A large choice of metals, bridging halogens, ligand molecules, and
counter ions enables us to investigate electron-phonon cooperative
phenomena in the one-dimensional Peierls-Hubbard system
systematically \cite{O2023}.
\widetext
\begin{figure}
\centerline
{\quad\mbox{\psfig{figure=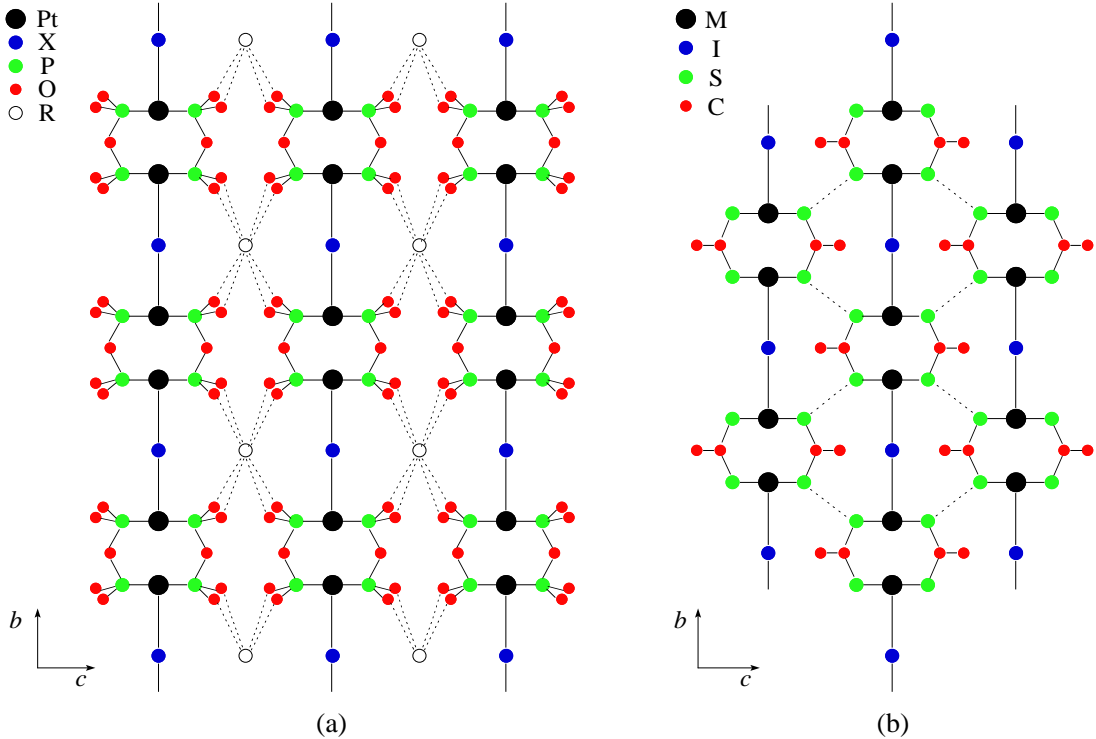,width=160mm,angle=0}}}
\vskip 4mm
\caption{Crystal structures of the pop complexes
         $R_4$[Pt$_2$(pop)$_4X$]$\cdot$$n$H$_2$O (a) and the dta
         complexes $M_2$(dta)$_4$I (b) in the projection of the
         $bc$ plane, where hydrogen atoms are omitted.
         The dotted lines represent hydrogen bonds in (a), while they
         correspond to van der Waals contacts in (b).}
\label{F:cryst}
\end{figure}
\narrowtext

\begin{figure}
\centerline
{\mbox{\psfig{figure=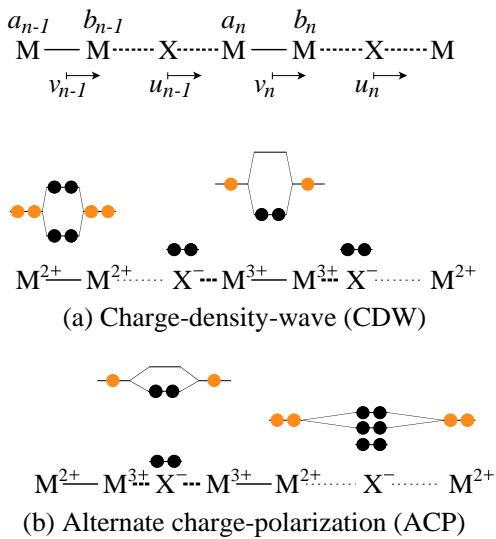,width=110mm,angle=0}}}
\vskip 4mm
\caption{Schematic representation of the two distinct ground states
         of $M\!M\!X$ chains [19]:
         (a) Charge-density-wave (CDW) state whose $X$ sublattice is
         dimerized;
         (b) Alternate charge-polarization (ACP) state whose $M_2$
         sublattice is dimerized.}
\label{F:CDW}
\end{figure}

   In recent years, binuclear metal analogs ($M\!M\!X$ chains) have
made our argument more and more exciting.
They comprise two families:
 $R_4$[Pt$_2$(pop)$_4X$]$\cdot$$n$H$_2$O
($X=\mbox{Cl},\mbox{Br},\mbox{I}$;
 $R=\mbox{Li},\mbox{K},\mbox{Cs},\cdots$;
 $\mbox{pop}=\mbox{diphosphonate}
 =\mbox{P}_2\mbox{O}_5\mbox{H}_2^{\,2-}$)
 \cite{C4604,C409} [Fig. \ref{F:cryst}(a)] and
 $M_2$(dta)$_4$I
($M=\mbox{Pt},\mbox{Ni}$;
 $\mbox{dta}=\mbox{dithioacetate}=\mbox{CH}_3\mbox{CS}_2^{\,-}$)
 \cite{B444,B2815} [Fig. \ref{F:cryst}(b)].
The pop complexes are in general semiconductors with
charge-density-wave ground states of the conventional type
\cite{B1155,K40} [Fig. \ref{F:CDW}(a)].
However, due to the small Peierls gaps, their ground states can be
tuned by replacing the halogens and/or counter ions \cite{Y2321}.
Such tuning of the electronic state is feasible also by pressure
\cite{S1405}.
On the other hand, the dta-family platinum complex, Pt$_2$(dta)$_4$I,
exhibits metallic conduction at room temperature \cite{K1931}, which
has never been realized in conventional $M\!X$ compounds.
With decreasing temperature, there occurs a metal-semiconductor
transition at $300$ K and further transition to a novel
charge-ordering mode [Fig. \ref{F:CDW}(b)] follows around $80$ K
\cite{K10068}.

   The exciting observations have stimulated theoretical research.
Carrying out quantum-chemical band calculations, Borshch {\it et al.}
\cite{B4562} attributed the novel valence distribution in
Pt$_2$(dta)$_4$I to its structural distortion.
Kuwabara and Yonemitsu \cite{K1545} more extensively investigated the
charge ordering and lattice modulation using various numerical tools.
The author \cite{Y183} also studied the ground-state properties with
particular emphasis on the contrast between the dta and pop
complexes.
Quantum, thermal, and pressure-induced phase transitions
were theoretically demonstrated \cite{Y1198,Y140102} so as to
interpret the observations, while the optical conductivity was
calculated \cite{K947} in an attempt to evaluate the hopping
amplitudes and Coulomb interactions.
\vspace*{-32mm}

   Now we are led to take a step toward the excited states.
Solitonic excitations are expected of doubly (or multiply in general)
degenerate charge-density-wave systems.
Ichinose \cite{I137} introduced the idea of domain walls into $M\!X$
chains and Onodera \cite{O250} developed the argument pointing out
some similarities between $M\!X$ chains and the trans isomer of
polyacetylene in the weak-coupling region.
Baeriswyl and Bishop \cite{B339} extended the calculation to
polaronic states laying emphasis on the strong-coupling region.
Optical absorption spectra due to solitons and polarons
\cite{G6408,T4074} and their relaxation process \cite{M5758,S1605}
in $M\!X$ chains were extensively investigated.
Although photoexperiments on $M\!M\!X$ chains \cite{Mats} are still
in their early stage, the optical excitations must be a coming
potential subject.
Making full use of a simple but relevant electron-phonon model, we
give a detailed description of intrinsic defects in $M\!M\!X$ chains.
Numerical calculations, together with an analytic argument in the
weak-coupling region, reveal that $M\!M\!X$ chains still exhibit a
striking analogy with {\it trans}-polyacetylene in their excitation
mechanism.
\begin{figure}
\vspace*{-14mm}
\centerline
{\mbox{\psfig{figure=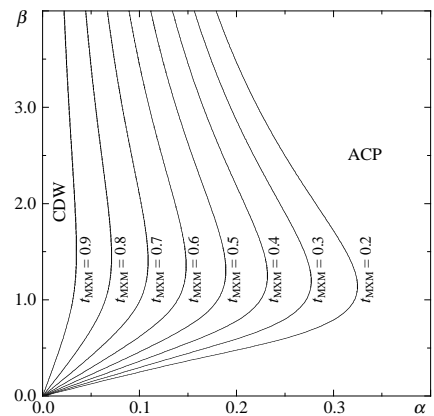,width=70mm,angle=0}}\quad}
\vskip -24mm
\caption{The $\alpha$-versus-$\beta$ ground-state phase diagrams at
         various values of $t_{\rm MXM}$.}
\label{F:PhD}
\end{figure}

\section{Model Hamiltonian}

   We employ the $\frac{3}{4}$-filled one-dimensional single-band
two-orbital electron-phonon model:
\begin{eqnarray}
   &&
   {\cal H}
    = -t_{\rm MM}\sum_{n,s}
       (a_{n,s}^\dagger b_{n,s}+b_{n,s}^\dagger a_{n,s})
   \nonumber \\
   &&\quad
    - \sum_{n,s}
       \bigl[
        t_{\rm MXM}-\alpha(l_{n+1}^{\rm (a)}+l_n^{\rm (b)})
       \bigr]
       (a_{n+1,s}^\dagger b_{n,s}+b_{n,s}^\dagger a_{n+1,s})
   \nonumber \\
   &&\quad
    - \beta\sum_{n,s}
       (l_n^{\rm (a)}a_{n,s}^\dagger a_{n,s}
       +l_n^{\rm (b)}b_{n,s}^\dagger b_{n,s})
   \nonumber \\
   &&\quad
    + \frac{K_{\rm MX}}{2}\sum_{n}
       \bigl[
        (l_n^{\rm (a)})^2+(l_n^{\rm (b)})^2
       \bigr]\,,
   \label{E:H}
\end{eqnarray}
where $a_{n,s}^\dagger$ and $b_{n,s}^\dagger$ are the creation
operators of an electron with spin $s=\pm$ (up and down) for the $M$
$d_{z^2}$ orbitals in the $n$th $M\!M\!X$ unit.
$t_{\rm MM}$ and $t_{\rm MXM}$ describe the intradimer and interdimer
electron hoppings, respectively.
$\alpha$ and $\beta$ are the intersite and intrasite electron-phonon
coupling constants, respectively, with $K_{\rm MX}$ being the
metal-halogen spring constant.
$l_n^{\rm (a)}=v_n-u_{n-1}$ and $l_n^{\rm (b)}=u_n-v_n$ with $u_n$
and $v_n$ being, respectively, the chain-direction displacements of
the halogen and metal dimer in the $n$th $M\!M\!X$ unit from their
equilibrium position.
We assume, based on the thus-far reported experimental observations,
that every $M_2$ moiety is not deformed.
The notation is further explained in Fig. \ref{F:CDW}.
We set $t_{\rm MM}$ and $K$ both equal to unity in the following.

   The Hamiltonian (\ref{E:H}) convincingly describes the two
distinct ground states of $M\!M\!X$ chains, which are schematically
shown in Fig. \ref{F:CDW}.
The CDW and ACP states are characterized by the alternating on-site
electron affinities and interdimer transfer energies, respectively.
Orbital hybridization mainly stays within each $M_2$ moiety in the
valence-trapped CDW state, while it essentially extends over
neighboring $M_2$ moieties in the valence-delocalized ACP state.
Therefore, the CDW state is more stabilized by increasing $\beta$ and
$t_{\rm MM}$, whereas the ACP state by increasing $\alpha$ and
$t_{\rm MXM}$.
The $M_2$ moieties are tightly locked together in the pop complexes
due to the hydrogen bonds between the ligands and counter cations,
while they are rather movable in the dta complexes owing to the
neutral chain structure.
Thus a significantly larger $\alpha$ is expected for the dta
complexes.
In this context, Fig. \ref{F:PhD} is well consistent with
experimental observations; (NH$_4$)$_4$[Pt$_2$(pop)$_4$Cl] exhibits a
ground state of the CDW type \cite{K40}, while Pt$_2$(dta)$_4$I
displays that of the ACP type \cite{K10068}.
In the weak-coupling region, $v_n=0$ in CDW, whereas $u_n=0$ in ACP.
With increasing coupling constants, their arise {\it mixed} states
with $u_n\neq 0$ and $v_n\neq 0$ \cite{B339,K2653,S479}.
We note that the present CDW and ACP are characterized as the lattice
configurations $l_n^{(a)}=l_n^{(b)}$ and $l_n^{(a)}=-l_n^{(b)}$,
respectively, regardless of each displacement $u_n$ and $v_n$.

\section{Results and Discussion}
\subsection{Variational Treatment}

   We consider local excitations from these ground states.
Under the constraint of the total chain length being unchanged, trial
wave functions may be introduced as
\begin{equation}
   l_n^{\rm (a)}=\sigma l_n^{\rm (b)}
  =(-1)^n l_0{\rm tanh}\bigl[(na-x_0)/\xi\bigr]\,,
   \label{E:WFS}
\end{equation}
for solitons \cite{S1698} and as
\begin{eqnarray}
   &&
   l_n^{\rm (a)}=\sigma l_n^{\rm (b)}
  =(-1)^n l_0
   \bigl\{
    1-{\rm tanh}(2\kappa d)
    \nonumber \\
   &&\quad\times
    \bigl[
     {\rm tanh}\kappa(na-x_0+d)
    -{\rm tanh}\kappa(na-x_0-d)
    \bigr]
   \bigr\}\,,
   \label{E:WFP}
\end{eqnarray}
for polarons \cite{B4,C4859}, where $\sigma$ takes $+$ and $-$ on the
CDW and ACP backgrounds, respectively, $a$ is the lattice constant,
and $l_0$ is the metal-halogen bond-length change in the ground
state.
$x_0$ indicates the defect center in both functions, while $\xi$
and $\kappa^{-1}{\rm tanh}(2\kappa d)$ describe the spatial extents of
solitons and polarons, respectively, all of which are variationally
determined.
Since we impose the periodic boundary condition on the Hamiltonian,
the soliton solutions assume that the number of the original unit
cells, $N$, should be odd, whereas the polaron solutions assume that
$N$ should be even.
We set $N$ equal to $201$ or $200$, either of which is much larger
than any defect extent in our calculation.
Defect energies are degenerate with respect to their spin, but it is
not the case with their charge due to the broken electron-hole
symmetry.
When we compare the defects on the CDW and ACP backgrounds, we
calculate them near the phase boundary so as to illuminate their
essential differences.
Taking the structural analyses \cite{C4604,C409,B444} into
consideration, we set $t_{\rm MM}=2t_{\rm MXM}$.

\subsection{Spatial Configuration and Energy Structure}

\begin{figure}
\vspace*{-58mm}
\centerline
{\mbox{\qquad\psfig{figure=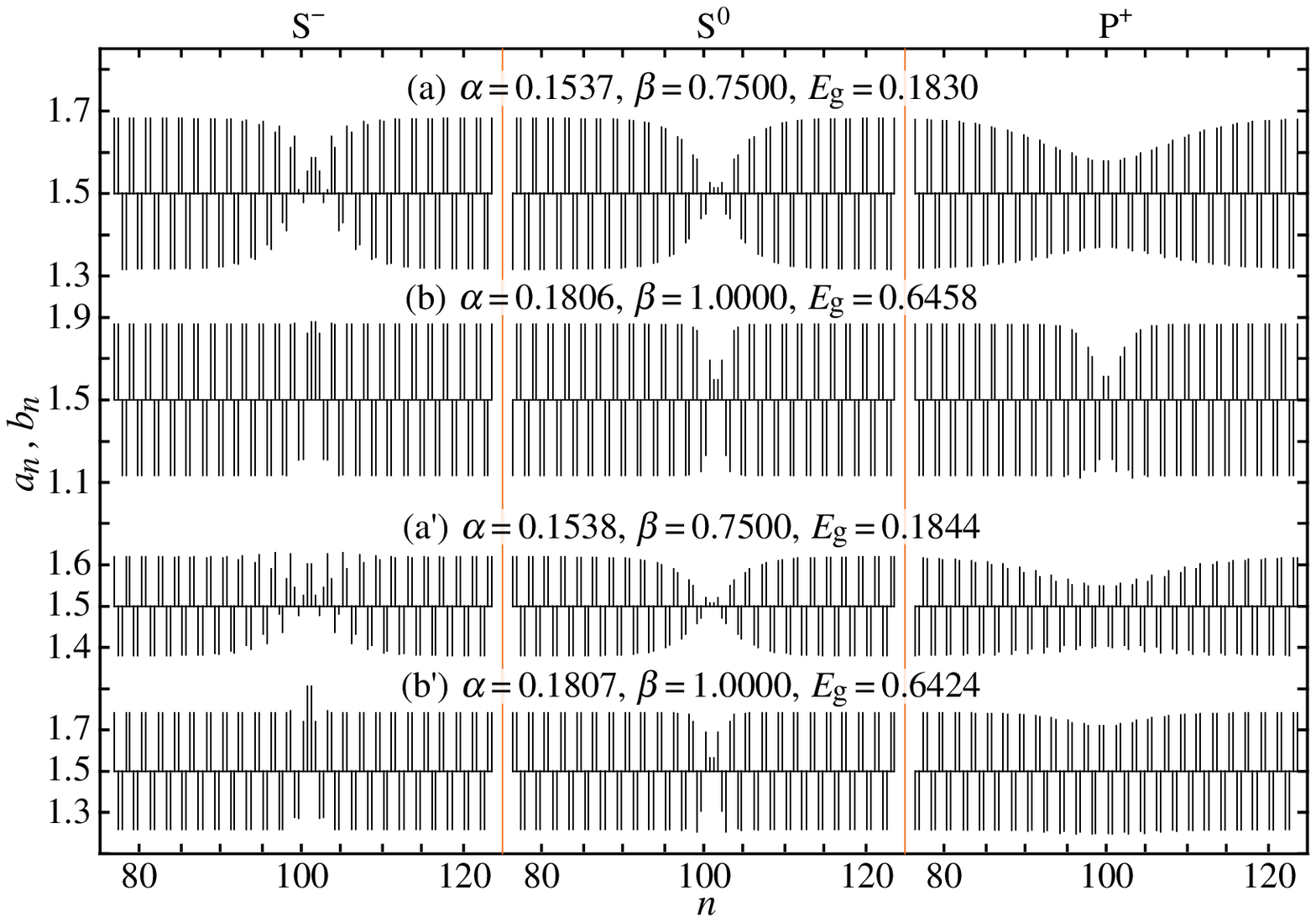,width=140mm,angle=0}}\qquad}
\vspace*{2mm}
\caption{Spatial configurations of negatively charged solitons S$^-$,
         neutral solitons S$^0$, and positively charged polarons
         P$^+$ in the CDW (a and b) and ACP (a$'$ and b$'$) states,
         where quantum averages of the local electron densities,
         $\sum_s\langle a_{n,s}^\dagger a_{n,s}\rangle\equiv a_n$ and
         $\sum_s\langle b_{n,s}^\dagger b_{n,s}\rangle\equiv b_n$,
         are measured in comparison with the average occupancy.}
\label{F:ED}
\vspace*{-2mm}
\centerline
{\quad\mbox{\psfig{figure=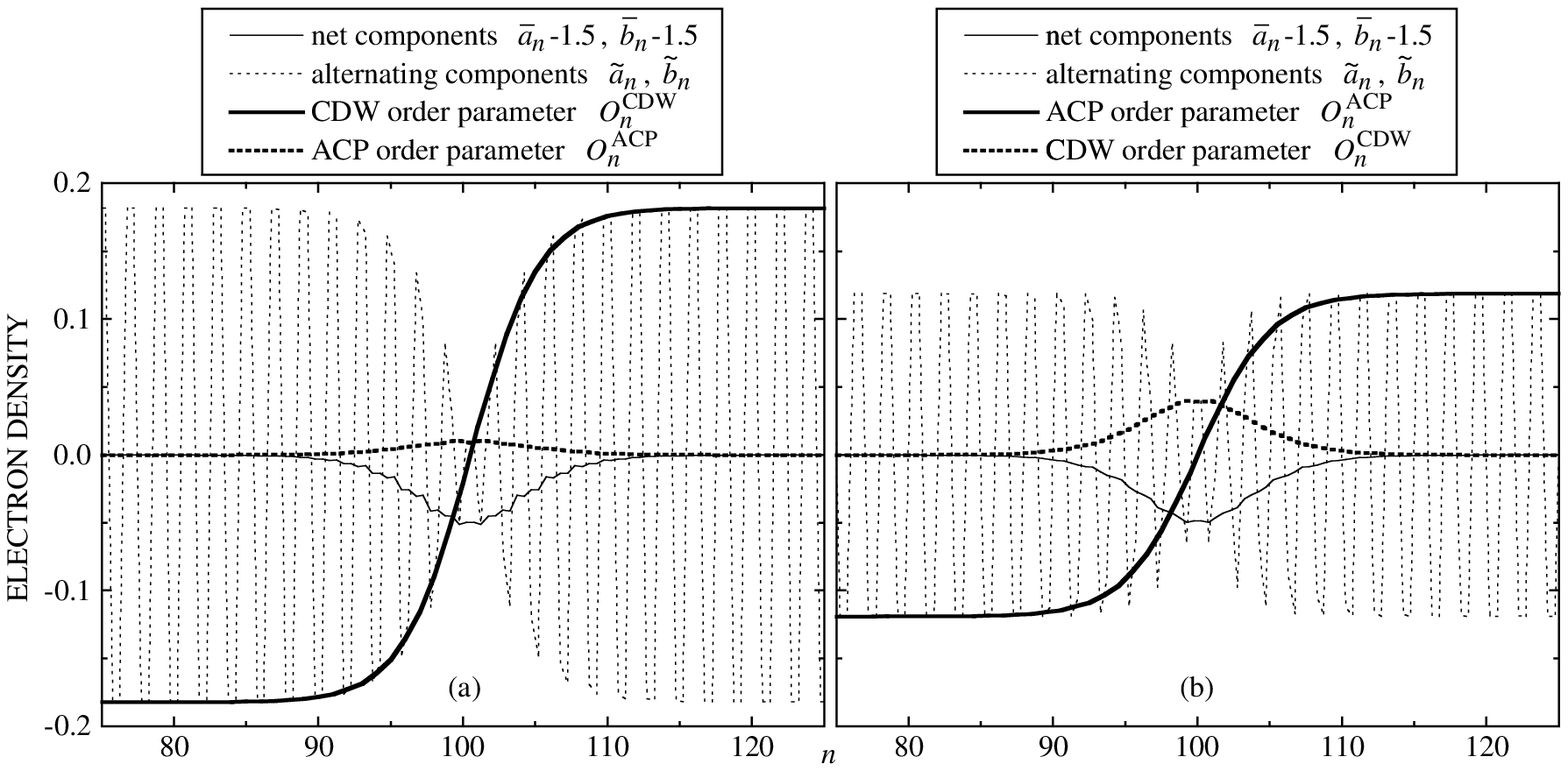,width=132mm,angle=0}}}
\vspace*{-14mm}
\caption{Spatial configurations of positively charged solitons S$^+$
         on the
         CDW (a; $\alpha=0.1537$, $\beta=0.7500$) and
         ACP (b; $\alpha=0.1538$, $\beta=0.7500$) backgrounds,
         where their alternating and net components (see in the text)
         are separately shown and the CDW and ACP order parameters
         (see in the text) are also plotted.}
\label{F:EDdec}
\end{figure}

   We show in Fig. \ref{F:ED} the spatial configurations of the
optimum defect solutions.
Spin and charge are separate entities in the solitonic excitations,
while the polaronic excitations exhibit the conventional spin-charge
characteristics.
Neutral solitons S$^0$ and polarons P$^\pm$ convey net spin densities
$\pm\frac{1}{2}$, whereas no spin density accompanies charged
solitons S$^\pm$.
Their formation energies do not depend on their locations in the
weak-coupling region, but the degeneracy is lifted with increasing
coupling strength (Table \ref{T:x0}).
As the Peierls gap increases, defects generally possess increasing
energies and decreasing extents, and end up with immobile entities.

   In order to investigate their electronic structures in more
detail, we decompose the electron densities $a_n$ and $b_n$ into
alternating ($\widetilde{a}_n$, $\widetilde{b}_n$) and net
(nonalternating) ($\overline{a}_n$, $\overline{b}_n$) components as
\begin{equation}
   \left.
   \begin{array}{ll}
    a_n=\overline{a}_n+\widetilde{a}_n\,;\ \ &
    \overline{a}_n=\frac{\displaystyle 1}{\displaystyle 4}
              (a_{n-1}+2a_n+a_{n+1})\,,\\
    b_n=\overline{b}_n+\widetilde{b}_n\,;\ \ &
    \overline{b}_n=\frac{\displaystyle 1}{\displaystyle 4}
              (b_{n-1}+2b_n+b_{n+1})\,,\\
   \end{array}
   \right.
\end{equation}
and show them for positively charged solitons S$^+$ in Fig.
\ref{F:EDdec}.
Now we can easily find out the positive charges accompanying the
defects.
One of the most interesting and important features of local
excitations of this kind is their interfacial character.
Let us define further the CDW and ACP order parameters using the
alternating components as
\begin{equation}
   \left.
   \begin{array}{lll}
    O_n^{\rm CDW}
    &=&\frac{\displaystyle (-1)^n}{\displaystyle 2}
       \left(
        \widetilde{a}_n+\widetilde{b}_n
       \right)\,,\\
    O_n^{\rm ACP}
    &=&\frac{\displaystyle (-1)^n}{\displaystyle 2}
       \left(
        \widetilde{a}_n+\widetilde{b}_{n-1}
       \right)\,,\\
   \end{array}
   \right.
\end{equation}
which detect oscillations of the CDW and ACP types, respectively,
and are also plotted in Fig. \ref{F:EDdec}.
We learn that the CDW soliton reduces the CDW amplitude and induces
the ACP amplitude instead in its center, while vice versa for the
ACP solion.
Their trajectories are expressed as
${\rm CDW}\rightarrow{\rm ACP}\rightarrow\overline{\rm CDW}$ and
${\rm ACP}\rightarrow\overline{\rm CDW}\rightarrow\overline{\rm ACP}$,
where $\overline{\rm A}$ denotes the antiphase counterpart for the
state A.
Neutral solitons and polarons exhibit different types of interface,
where certain kinds of spin-density-wave oscillation blend into CDW
or ACP.
Due to their mixed nature, solitons and polarons could be divided
into more elementary defects in the vicinity of the phase boundaries.

\begin{figure}
\vspace*{-60mm}
\centerline
{\mbox{\psfig{figure=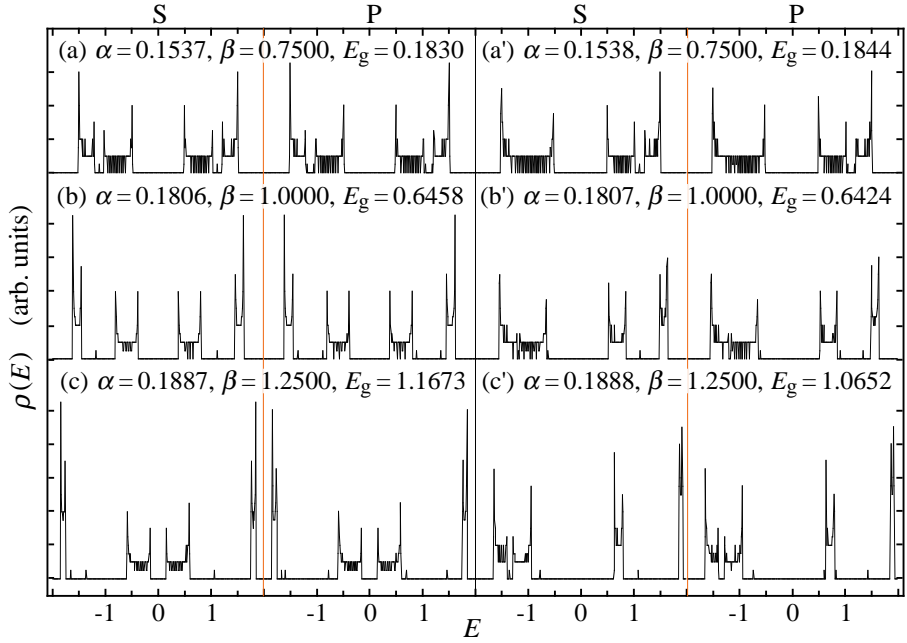,width=142mm,angle=0}}}
\vspace*{-60mm}
\caption{Density of states $\rho(E)$ for the optimum soliton and
         polaron solutions on the
         CDW ($\mbox{a},\mbox{b},\mbox{c}$) and
         ACP ($\mbox{a}',\mbox{b}',\mbox{c}'$) backgrounds.}
\label{F:DOS}
\end{figure}

   Density of states for solitons and polarons are shown in Fig.
\ref{F:DOS}.
The four bands coming from the CDW background are, from the bottom to
the top, largely made up of the bonding combination $\phi_+$ of
binuclear Pt$^{2+}$-Pt$^{2+}$ units, that of Pt$^{3+}$-Pt$^{3+}$
units, the antibonding combination $\phi_-$ of Pt$^{2+}$-Pt$^{2+}$
units, and that of Pt$^{3+}$-Pt$^{3+}$ units.
On the other hand, the major components of the four bands coming from
the ACP background are
the $\phi_+$ orbitals of interdimer Pt$^{2+}$-X$^-$-Pt$^{2+}$ units,
the $\phi_-$ orbitals of Pt$^{2+}$-X$^-$-Pt$^{2+}$ units,
the $\phi_+$ orbitals of Pt$^{3+}$-X$^-$-Pt$^{3+}$ units, and
the $\phi_-$ orbitals of Pt$^{3+}$-X$^-$-Pt$^{3+}$ units.
Therefore, increasing $\beta$ splits both $\sigma$ and $\sigma^*$
orbitals in the CDW state, while in the ACP state the splitting
between $\sigma$ and $\sigma^*$ orbitals is stressed as an effect of
increasing $\beta$.
\widetext

\narrowtext

   The soliton solutions commonly exhibit an additional level within
the gap, whereas the polaron ones are accompanied by a couple of
intragap levels.
These levels are all found to be localized around the defect center.
The level structures in the weak-coupling region suggest some
similarities between small-gap $M\!M\!X$ chains and
{\it trans}-polyacetylene.
However, in contrast with the case of polyacetylene
\cite{S1698,C4859}, the level structure which is symmetric with
respect to the middle of the gap greaks down with increasing
electron-phonon interactions due to the lack of electron-hole
symmetry.
Besides the intragap levels, there appear a few related levels which
are also localized around the defect center.

\subsection{Scaling Properties}

\begin{figure}
\centerline
{\mbox{\psfig{figure=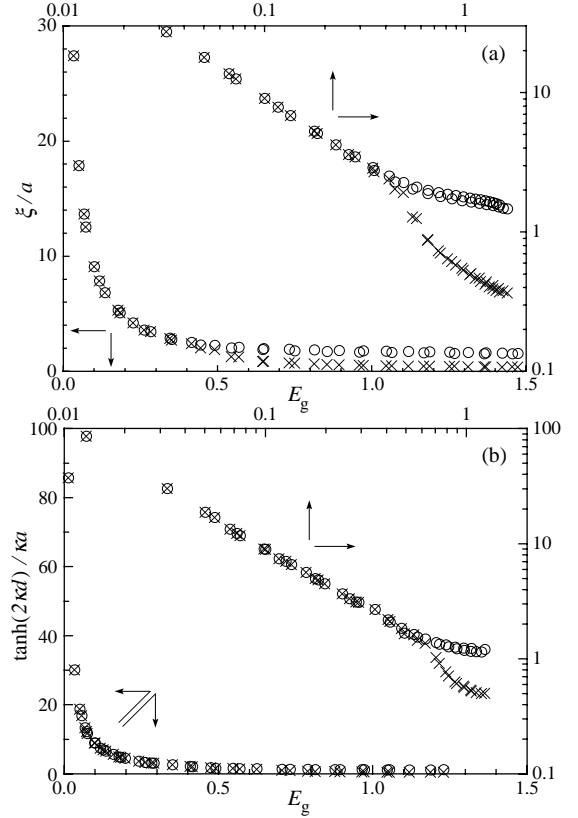,width=74mm,angle=0}}}
\vskip 2mm
\caption{The optimized extents of the solitons (a) and polarons (b)
         as functions of the band gap $E_{\rm g}$ in linear and
         logarithmic scales under various values of $\alpha$ and
         $\beta$, where $\circ$ and $\times$ correspond to the
         highest- and lowest-energy locations $x_0$, respectively.}
\label{F:xi}
\end{figure}

\begin{figure}
\vspace*{-40mm}
\centerline
{$\!\!\!\!\!\!\!\!$
 \mbox{\psfig{figure=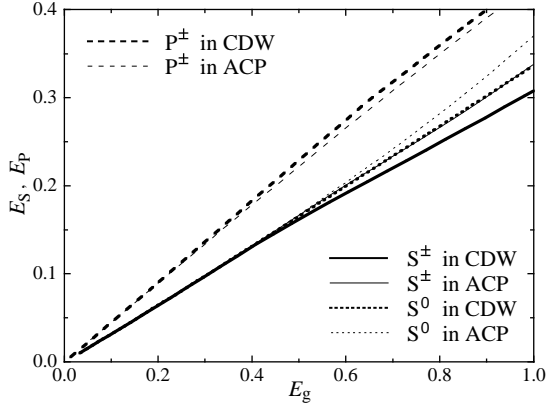,width=92mm,angle=0}}\quad}
\vskip -50mm
\caption{The soliton and polaron formation energies averaged over
         $x_0$ as functions of the band gap $E_{\rm g}$.
         The negatively (positively) charged solitons and polarons
         possess more (less) than three-quarter-filled electron
         bands and therefore their formation energies can not
         be defined in themselves.
         As for charged excitations, $E_{\rm S}$ ($E_{\rm P}$) is
         further averaged over S$^+$ (P$^+$) and S$^-$ (P$^-$).}
\label{F:Ef}
\end{figure}

   In order to illuminate the scaling properties of the solitons and
polarons in more detail, we plot in Fig. \ref{F:xi} their
characteristic lengths $\xi$ and $\kappa^{-1}{\rm tanh}(2\kappa d)$
as functions of the band gap $E_{\rm g}$.
Although we have calculated them with varying $\alpha$, $\beta$, and
$x_0$, they are uniquely scaled by $E_{\rm g}$ in the weak-coupling
region as
\begin{equation}
   \xi/a\sim 0.95/E_{\rm g}^{0.98}\,,\ \
   {\rm tanh}(2\kappa d)/\kappa a
   \sim 0.88/E_{\rm g}^{1.03}\,,
   \label{E:xi}
\end{equation}
where the second decimal place varies with the data set taken into
the fitting.
It is convincing that such scaling laws break down as the defect
extents come close to the length scale $a$.
The formation energies are also scaled by the gap, as is shown in
Fig. \ref{F:Ef}.
In the weak-coupling region, the soliton and polaron energies are
both degenerate with respect to their spin and location, and exhibit
scaling formulae

\vskip 2mm

\begin{equation}
   E_{\rm S}\sim 0.33E_{\rm g}\,,\ \ 
   E_{\rm P}\sim 0.46E_{\rm g}\,.
   \label{E:Ef}
\end{equation}
With increasing gap, the degeneracy is lifted, where the CDW solitons
have lower energies than their ACP counterparts at a given
$E_{\rm g}$, while vice versa for polarons.
Now that we have obtained the numerical observations (\ref{E:Ef}), we
should be reminded of the defect states in {\it trans}-polyacetylene
and a rigorous approach \cite{C4859,T2388} to them.

   Su, Schrieffer, and Heeger (SSH) \cite{S1698} pioneeringly
presented a theoretical study on the soliton excitations in
polyacetylene.
Although they employed a simple electron-phonon model, essential
features of solitons such as the midgap localized energy level and
the small effective mass were successfully interpreted.
Takayama, Lin-Liu, and Maki (TLM) \cite{T2388} mapped the SSH model
onto a continuous line and obtained a rigorous soliton solution with
the formation energy $E_{\rm S}=E_{\rm g}/\pi$.
A polaron solution \cite{B4,C4859} was also found within the same
scheme and its formation energy was revealed as
$E_{\rm P}=\sqrt{2}E_{\rm g}/\pi$.
Our numerical estimates (\ref{E:Ef}), which are very close to the TLM
findings, suggest that $M\!M\!X$ chains still have an analogy with
polyacetylene.
Inquiring the continuum-model description of $M\!M\!X$ chains, we try
to understand the numerical calculations.

   The continuum version of the present model (\ref{E:H}) should be
constructed from the primary $d_{\sigma^*}$ conduction band under no
dimerization of the system.
Without any electron-phonon coupling, the Hamiltonian (\ref{E:H}) can
be expressed as
\begin{equation}
   \lim_{\alpha,\beta\rightarrow 0}{\cal H}
   =\sum_{k,s}
    \left(
     \varepsilon_k^+ A_{k,s}^\dagger A_{k,s}
    +\varepsilon_k^- B_{k,s}^\dagger B_{k,s}
    \right)\,,
\end{equation}
where
\begin{equation}
   \varepsilon_k^\pm
   =\pm\sqrt{t_{\rm MM}^2+t_{\rm MXM}^2
            +2t_{\rm MM}t_{\rm MXM}{\rm cos}ka}\,,
\end{equation}
\begin{equation}
   \left.
   \begin{array}{lll}
    A_{k,s}
    &=&\frac{\displaystyle 1}{\displaystyle\sqrt{2}}
       \left(
        {\rm e}^{{\rm  i}\theta/2}a_{k,s}
       +{\rm e}^{{\rm -i}\theta/2}b_{k,s}
       \right)\,,\\
    B_{k,s}
    &=&\frac{\displaystyle 1}{\displaystyle\sqrt{2}}
       \left(
        {\rm e}^{ {\rm i}\theta/2}a_{k,s}
       -{\rm e}^{-{\rm i}\theta/2}b_{k,s}
       \right)\,,\\
   \end{array}
   \right.
\end{equation}
with $a_{k,s}$ ($b_{k,s}$) being the Fourier transform of $a_{n,s}$
($b_{n,s}$) and
\begin{equation}
   {\rm e}^{{\rm i}\theta}
   =-\frac{t_{\rm MM}+t_{\rm MXM}{\rm e}^{{\rm i}ka}}
          {\sqrt{t_{\rm MM}^2+t_{\rm MXM}^2
                +2t_{\rm MM}t_{\rm MXM}{\rm cos}ka}}\,.
\end{equation}
As far as we restrict our interest to the low-lying excitations, we
may discard the irrelevant $d_\sigma$ band, which is here described
by the dispersion relation $\varepsilon_k^-$, and linearize the
relevant dispersion $\varepsilon_k^+$ at the two Fermi points.
Taking account of the electron-phonon coupling within this scheme and
assuming the CDW and ACP backgrounds, we obtain two effective 
Hamiltonians in the coordinate representation:
\begin{eqnarray}
   &&
   {\cal H}_{\rm eff}^{\rm CDW}
  =t_{\rm eff}\sum_{n,s}
   \left[
    1-\frac{{\mit\Delta}(na)}{t_{\rm MM}}
   \right]
   \left(
    A_{n,s}^\dagger A_{n+1,s}+{\rm H.c.}
   \right)
   \nonumber \\
   &&\ 
  +\frac{2t_{\rm eff}}{t_{\rm MXM}}\sum_{n,s}
   {\mit\Delta}(na)
   A_{n,s}^\dagger A_{n,s}
  +\frac{K_{\rm MX}}{4\beta_{\rm eff}^2}\sum_n
   {\mit\Delta}(na)^2\,,
   \\
   &&
   {\cal H}_{\rm eff}^{\rm ACP}
  =t_{\rm eff}\sum_{n,s}
   \left[
    1-\frac{{\mit\Delta}(na)}{t_{\rm MXM}}
   \right]
   \left(
    A_{n,s}^\dagger A_{n+1,s}+{\rm H.c.}
   \right)
   \nonumber \\
   &&\ 
  +\frac{2t_{\rm eff}}{t_{\rm MM}}\sum_{n,s}
   {\mit\Delta}(na)
   A_{n,s}^\dagger A_{n,s}
  +\frac{K_{\rm MX}}{16\alpha_{\rm eff}^2}\sum_n
   {\mit\Delta}(na)^2\,,
\end{eqnarray}
where
\begin{equation}
   t_{\rm eff}=\frac{t_{\rm MM}t_{\rm MXM}}
                    {2\sqrt{t_{\rm MM}^2+t_{\rm MXM}^2}}\,,
   \label{E:teff}
\end{equation}
\begin{equation}
   \alpha_{\rm eff}
   =\frac{1}{2}
    \left(
     \alpha+\frac{t_{\rm eff}}{t_{\rm MM}}\beta
    \right)\,,\ \ 
   \beta_{\rm eff}
   =\frac{t_{\rm eff}}{t_{\rm MXM}}\beta\,,
\end{equation}
\begin{equation}
   {\mit\Delta}(na)
   =\left\{
     \begin{array}{l}
      2\beta_{\rm eff}(-1)^n(u_n-v_n)\ \ \ 
      {\rm for}\ {\cal H}_{\rm eff}^{\rm CDW}\,,\\
      4\alpha_{\rm eff}(-1)^n(u_n-v_n)\ \ \ 
      {\rm for}\ {\cal H}_{\rm eff}^{\rm ACP}\,.\\
     \end{array}
    \right.
\end{equation}
In accordance with the slowly-varying gap parameter
${\mit\Delta}(na)$, we introduce operators
\begin{equation}
   \left.
   \begin{array}{lll}
    \psi_s^{\rm (r)}(na)
    &=&\frac{\displaystyle 1}{\displaystyle\sqrt{L}}
       {\displaystyle\sum_k}
       {\rm e}^{{\rm i}kna}A_{ k_{\rm F}+k,s}\,,\\
    \psi_s^{\rm (l)}(na)
    &=&\frac{\displaystyle -{\rm i}}{\displaystyle\sqrt{L}}
       {\displaystyle\sum_k}
       {\rm e}^{{\rm i}kna}A_{-k_{\rm F}+k,s}\,,\\
   \end{array}
   \right.
\end{equation}
for right- and left-moving electrons, respectively, where
$L=Na$, $k_{\rm F}=\pi/2a$ and $-\pi/2a<k\leq\pi/2a$.
Now, summing out fast-varying components and taking the
$a\rightarrow 0$ limit, we reach the relevant continuum models
\begin{eqnarray}
   &&
   {\cal H}_{\rm eff}^{\rm CDW}
  =\int
   \frac{{\mit\Delta}(x)^2}{2\pi\hbar v_{\rm F}\lambda}
   {\rm d}x
  +\sum_s\int
   {\mit\Phi}_s^\dagger(x)
   \biggl[
    {\rm i}\hbar v_{\rm F}\sigma_z\frac{{\rm d}}{{\rm d}x}
   \nonumber \\
   &&\quad
  -2{\mit\Delta}(x)t_{\rm eff}
    \left(
     \frac{\sigma_x}{t_{\rm MM}}+\frac{\sigma_y}{t_{\rm MXM}}
    \right)
   \biggr]
   {\mit\Phi}_s(x)
   {\rm d}x\,,
   \label{E:HCDWc}
   \\
   &&
   {\cal H}_{\rm eff}^{\rm ACP}
  =\int
   \frac{{\mit\Delta}(x)^2}{2\pi\hbar v_{\rm F}\lambda}
   {\rm d}x
  +\sum_s\int
   {\mit\Phi}_s^\dagger(x)
   \biggl[
    {\rm i}\hbar v_{\rm F}\sigma_z\frac{{\rm d}}{{\rm d}x}
   \nonumber \\
   &&\quad
  -2{\mit\Delta}(x)t_{\rm eff}
    \left(
     \frac{\sigma_x}{t_{\rm MXM}}+\frac{\sigma_y}{t_{\rm MM}}
    \right)
   \biggr]
   {\mit\Phi}_s(x)
   {\rm d}x\,,
   \label{E:HACPc}
\end{eqnarray}
where we have employed the Pauli matrices
$(\sigma_x,\sigma_y,\sigma_z)$ and a spinor notation
${\mit\Phi}_s^\dagger(x)
 =(\psi_s^{\rm (r)}(x)^*,\psi_s^{\rm (l)}(x)^*)$,
defining the Fermi velocity $v_{\rm F}$ and the dimensionless
coupling constant $\lambda$ as
\begin{equation}
   \hbar v_{\rm F}=2at_{\rm eff}\,,
   \label{E:vF}
\end{equation}
\begin{equation}
   \lambda
   =\left\{
     \begin{array}{l}
      \beta_{\rm eff}^2/
       \pi t_{\rm eff}K_{\rm MX}\ \ \ 
      {\rm for}\ {\cal H}_{\rm eff}^{\rm CDW}\,,\\
      4\alpha_{\rm eff}^2/
       \pi t_{\rm eff}K_{\rm MX}\ \ \ 
      {\rm for}\ {\cal H}_{\rm eff}^{\rm ACP}\,.\\
     \end{array}
    \right.
\end{equation}
Interestingly, when we rotate the spinor wave function around the $z$
axis by the angle
\begin{equation}
   \varphi
   =\left\{
     \begin{array}{l}
      {\rm arctan}(t_{\rm MM}/t_{\rm MXM})\ \ \ 
      {\rm for}\ {\cal H}_{\rm eff}^{\rm CDW}\,,\\
      {\rm arctan}(t_{\rm MXM}/t_{\rm MM})\ \ \ 
      {\rm for}\ {\cal H}_{\rm eff}^{\rm ACP}\,,\\
     \end{array}
    \right.
\end{equation}
both Hamiltonians (\ref{E:HCDWc}) and (\ref{E:HACPc}) turn into the
same expression
\begin{eqnarray}
   &&
   {\cal H}_{\rm TLM}
  =\int
   \frac{{\mit\Delta}(x)^2}{2\pi\hbar v_{\rm F}\lambda}
   {\rm d}x
  +\sum_s\int
   {\mit\Phi}_s^\dagger(x)
   \nonumber \\
   &&\quad
  \times
   \left[
    -{\rm i}\hbar v_{\rm F}\sigma_z\frac{{\rm d}}{{\rm d}x}
    +{\mit\Delta}(x)\sigma_x
   \right]
   {\mit\Phi}_s(x)
   {\rm d}x\,,
   \label{E:HTLM}
\end{eqnarray}
which is nothing but the TLM continuum model.
We know that the Hamiltonian (\ref{E:HTLM}) possesses exact solitonic
\cite{T2388,K4173} and polaronic \cite{B4,C4859} solutions
\begin{eqnarray}
   &&
   {\mit\Delta}(x)
   ={\mit\Delta}_0
    {\rm tanh}(x/\xi)\,,
   \\
   &&
   {\mit\Delta}(x)
   ={\mit\Delta}_0
   -\hbar v_{\rm F}\kappa
   \nonumber \\
   &&\quad\times
    \bigl[
     {\rm tanh}\kappa(x+d)
    -{\rm tanh}\kappa(x-d)
    \bigr]\,,
\end{eqnarray}
with their characteristic scaling relations
\begin{equation}
   \xi=\hbar v_{\rm F}/{\mit\Delta}_0\,,\ \ 
   E_{\rm S}=2{\mit\Delta}_0/\pi\,,
\end{equation}
\begin{equation}
   {\rm tanh}(2\kappa d)/\kappa
   =\hbar v_{\rm F}/{\mit\Delta}_0\,,\ \ 
   E_{\rm P}=2\sqrt{2}{\mit\Delta}_0/\pi\,.
\end{equation}
Considering Eqs. (\ref{E:teff}) and (\ref{E:vF}) together with the
relation $2{\mit\Delta}_0=E_{\rm g}$, we are fully convinced of the
present numerical findings (\ref{E:xi}) and (\ref{E:Ef}).

\quad
\vspace*{-42mm}
\begin{figure}
\centerline
{$\!\!\!\!$\mbox{\psfig{figure=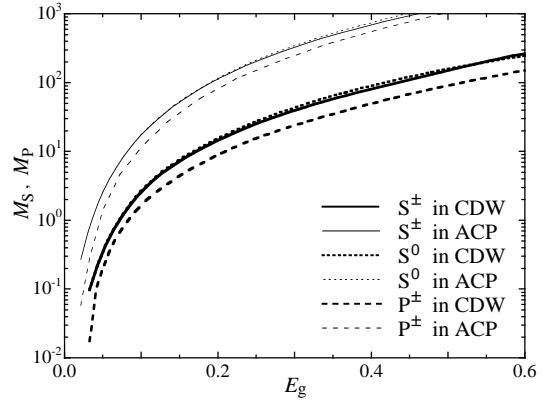,width=96mm,angle=0}}}
\vskip -48mm
\caption{The soliton and polaron masses as functions of the band gap
         $E_{\rm g}$ in the unit of the electron mass $m_{\rm e}$.
         All the excitations are almost degenerate with respect to
         their spin, charge, and location in the weak-coupling
         region, but the degeneracy is lifted with increasing gap.
         As for charged defects, we have simply plotted the masses of
         negatively charged ones, because those of positively charged
         ones are less convergent in the strong-coupling region, much
         more rapidly increasing with gap outside the weak-coupling
         region.}
\label{F:M}
\end{figure}

\subsection{Effective Mass}

   The continuum-model analysis has revealed that the defect extents
should be given by $2a/\sqrt{5}E_{\rm g}$ in the weak-coupling limit
under the present parametrization.
According to the numerical observations (\ref{E:xi}), this scaling
law better holds for polarons than for solitons.
This may be because polarons have wider extents than solitons at a
given gap.
When we consider moving defects, such an idea is quite suggestive.
Let us calculate the effective masses of solitons and polarons.

   Moving the defect center as $x_0=vt$ in Eqs. (\ref{E:WFS}) and
(\ref{E:WFP}) and neglecting any change in the defect shape, which
must be of order $v^2$ due to the time-reversal symmetry and
therefore does not contribute to the defect mass $M_{\rm defect}$ for
small $v$, we find
\begin{equation}
   \frac{1}{2}M_{\rm defect}v^2
   =\frac{1}{2}M\sum_n (\dot{l}_n^{\rm (a)})^2\,,
   \label{E:M}
\end{equation}
where $M$ corresponds to the mass of a halogen or bimetal complex
according as the ground state is CDW or ACP.
Assuming $n$ to be sufficiently large under the periodic boundary
condition, the summation in Eq. (\ref{E:M}) can explicitly be taken
and ends up with
\begin{eqnarray}
   &&
   M_{\rm S}
   =\frac{4Ma}{3\xi}
    \left(\frac{l_0}{a}\right)^2\,,
   \\
   &&
   M_{\rm P}
   =\frac{8}{3}M\kappa a
    \left(\frac{l_0}{a}\right)^2
    {\rm tanh}^2(2\kappa d)
   \nonumber \\
   &&\quad\times
    \biggl[
     1-12\frac{2\kappa d+1+(2\kappa d-1){\rm e}^{4\kappa d}}
              {{\rm e}^{12\kappa d}-3{\rm e}^{8\kappa d}
              +3{\rm e}^{8\kappa d}-1}
    \biggr]\,,
\end{eqnarray}
for solitons and polarons, respectively.
Further calculation is performed for two well-studied
$M\!M\!X$-chain materials
K$_4$[Pt$_2$(pop)$_4$Br]$\cdot$$3$H$_2$O
\cite{C4604,B1155,S1405,K4420,S66} with $a=8.139$ \AA \cite{B1155}
and Pt$_2$(dta)$_4$I
\cite{B444,K1931,K10068,S265,Miya} with $a=8.633$ \AA \cite{B444}.
Their ground states were assigned to CDW and ACP, respectively, and
therefore they give $M=145611m_{\rm e}$ and $M=1375629m_{\rm e}$,
respectively, where $m_{\rm e}$ is the electron mass.
We show in Fig. \ref{F:M} the thus calculated soliton and polaron
masses, where the optimum values of $\xi$, $\kappa$, and $d$ have
been averaged over $x_0$.
For $E_{\rm g}\agt 0.4$, any defect has to overcome site-dependent
energy barriers in its motion and thus the present calculations are
just for reference.

   In comparison with the formation energy which describes a static
defect and is generically given for both types of ground state in
the weak-coupling region, the effective mass of a moving defect
varies according to the background charge ordering.
Defects on the ACP background generally look more massive than
their CDW counterparts.
However, the difference should rather be attributed to the
material-dependent crystalline structure than be recognized as a
result of intrinsic defect features.
CDW is accompanied by the halogen-sublattice dimerization, while ACP
by the metal-sublattice dimerization.
Larger defect masses on the ACP background reflect the mass of the
diplatinum complex Pt$_2$(CH$_3$CS$_2$)$_4$ which is about ten times
as heavy as the halogen atom Br.

   On the other hand, the polaron mass turns out intrinsically
smaller than the soliton mass.
We have indeed found in Fig. \ref{F:ED} that the polarons are more
spatially extended than the solitons at a given gap.
The energy gap at least amounts to $0.75$ eV \cite{O9} in
conventional $M\!X$-chain materials.
Metal binucleation strikingly reduces the gap, for example, to $0.08$
eV for K$_4$[Pt$_2$(pop)$_4$Br]$\cdot$$3$H$_2$O \cite{B1155} and
$0.05$ eV for Pt$_2$(dta)$_4$I \cite{B444}.
Correspondingly, Pt$_2$(dta)$_4$I exhibits a high electrical
conductivity $13$ $\Omega^{-1}\mbox{cm}^{-1}$ \cite{K10068}, which is
about $10^9$ times larger than that of a typical $M\!X$ compound.
Though we have no information on $t_{\rm MM}$, previous estimates of
$t_{\rm MXM}$ for $M\!X$ compounds \cite{B239,T2212}, together with
the structural analyses \cite{C4604,C409,B444}, imply that
$t_{\rm MM}=1\sim 2$ eV for $M\!M\!X$ compounds.
Then we are allowed to estimate that
$0.1m_{\rm e}\alt M_{\rm S}\alt 2m_{\rm e}$ and
$0.05m_{\rm e}\alt M_{\rm P}\alt m_{\rm e}$ for
K$_4$[Pt$_2$(pop)$_4$Br]$\cdot$$3$H$_2$O, while
$0.3m_{\rm e}\alt M_{\rm S}\alt 3m_{\rm e}$ and
$0.1m_{\rm e}\alt M_{\rm P}\alt m_{\rm e}$ for Pt$_2$(dta)$_4$I.
In any case, the $M\!M\!X$-solitons are a few times as heavy as
the $M\!M\!X$-polarons but still much lighter than the
$M\!X$-solitons of $M_{\rm S}\agt 100m_{\rm e}$ \cite{O250}.
There is a similar observation in {\it trans}-polyacetylene as
well.
The TLM-model analysis of polyacetylene gives
$M_{\rm S}\simeq 3m_{\rm e}$ \cite{B23} and
$M_{\rm P}\simeq m_{\rm e}$ \cite{C4859}.
Such small masses should be quite significant in dynamics involving
solitons and polarons.
From the energetical point of view, we have learnt that
$E_{\rm P}<2E_{\rm S}<2E_{\rm P}$.
We can not well distinguish charged solitons from neutral ones within
the electron-phonon model, but we find, taking account of
electron-electron correlation, that $E_{{\rm S}^0}<E_{{\rm S}^\pm}$
\cite{W8546}.
Hence there may be a relaxation of photogenerated polaron pairs
$\mbox{P}^+ +\mbox{P}^-\longrightarrow\mbox{S}^0+\bar{\mbox{S}}^0$
and a recombination of doping-induced or chemical-defect-associated
solitons
$\mbox{S}^\pm+\bar{\mbox{S}}^0\longrightarrow\mbox{P}^\pm$.
In $M\!X$ compounds, photogeneration of solitons and polarons have
indeed been observed and they exhibit a varied relaxation process
within a few hundred psec \cite{K2122,O2248,O2023}.
$M\!M\!X$ chains with less massive defects may really exhibit
picosecond dynamics.

\section{Concluding Remarks}

   We have investigated the local excitations corresponding to
topological solitons and polarons in halogen-bridged diplatinum
chains in terms of a coupled electron-phonon model.
Although the solitons have lower excitation energies than the
polarons, the effective masses of the solitons are larger than those
of polarons.
Their high mobility is promised in $M\!M\!X$ chains with small
energy gaps up to $\sim 0.1$ eV, where all the defects are at most a
few times as massive as an electron.

   There is an analogy between the defect states in $M\!M\!X$ chains
and those in {\it trans}-polyacetylene in spite of many contrasts
between these systems such as $d$ and $\pi$ electrons, binuclear and
mononuclear assemblies, and intrasite and intersite electron-phonon
couplings.
Then we are all the more eager to encourage optical measurements and
doping of $M\!M\!X$ chains, where the electronic gap is suitably
small, chemical tuning of the electronic state is widely possible,
and the single crystals are readily available.

\acknowledgments

   The author is grateful to A. R. Bishop, K. Yonemitsu, K. Nasu,
and Y. Ono for fruitful discussions and helpful comments.
This work was supported by the Japanese Ministry of Education,
Science, and Culture and by the Sumitomo Foundation.
The numerical calculation was done using the facility of the
Supercomputer Center, Institute for Solid State Physics, University
of Tokyo.

\begin{table}
\caption{The lowest- and highest-energy locations $x_0$ of $M\!M\!X$
         solitons and polarons.}
\begin{tabular}{lcccc}
&
\multicolumn{2}{c}{\ CDW background\ } &
\multicolumn{2}{c}{\ ACP background\ } \\
\cline{2-3}
\cline{4-5}
&
the lowest & the highest & the lowest & the highest \\
\tableline
\noalign{\vskip 2pt}
\ S$^-$ & $(2n-\frac{1}{2})a$ & $(2n+\frac{1}{2})a$ &
          $(2n+1)a$           & $2na$               \\
\ S$^+$ & $(2n+\frac{1}{2})a$ & $(2n-\frac{1}{2})a$ &
          $2na$               & $(2n+1)a$           \\
\ S$^0$ & $(2n-\frac{1}{2})a$ & $(2n+\frac{1}{2})a$ &
          $(2n+1)a$           & $2na$               \\
\ P$^-$ & $(2n+1)a$           & $2na$               &
          $(2n+\frac{1}{2})a$ & $(2n-\frac{1}{2})a$ \\
\ P$^+$ & $2na$               & $(2n+1)a$           &
          $(2n-\frac{1}{2})a$ & $(2n+\frac{1}{2})a$ \\
\end{tabular}
\label{T:x0}
\end{table}

\widetext
\end{document}